\documentstyle[twocolumn,aps]{revtex}

\begin{document}

\title{A Note on ADE-Spectra in Conformal Field Theory}
\author{{\bf {\large A.G.~Bytsko $^{\dag}$ and A.~Fring $^*$}}}
\address{\noindent $^{\dag}$ Steklov Mathematics Institute,\\
Fontanka 27, St. Petersburg 191011, Russia \\
$^*$ Institut f\"ur Theoretische Physik,\\
Freie Universit\"at Berlin, Arnimallee 14, D-14195 Berlin, Germany}
\date{December, 1998}
\maketitle

\begin{abstract}
We demonstrate that certain Virasoro characters (and their linear 
combinations) in minimal and non-minimal conformal models which admit 
factorized forms are manifestly related to the ADE series. This permits 
to extract quasi-particle spectra of a Lie algebraic nature which 
resembles the features of Toda field theory. These spectra possibly 
admit a construction in terms of the $W_n$-generators. In the course of 
our analysis we establish interrelations between the factorized characters 
related to the parafermionic models, the compactified boson and the 
minimal models.
\end{abstract}

\section{Introduction}

\footnotetext{
e-mail addresses:
\par
\begin{tabular}{l}
bytsko@pdmi.ras.ru,  \\ 
fring@physik.fu-berlin.de. 
\end{tabular} }

It is well known, that a large class of off-critical integrable models is
related to affine Toda field theories \cite{Toda} or RSOS-statistical
models \cite{RSOS}, which posses a rich underlying Lie algebraic structure.
Since these models can be regarded as perturbed conformal field theories, 
it is suggestive to recover the underlying Lie algebraic structure also 
in the conformal limit. Of primary interest is  to identify the conformal
counterparts of the off-critical particle spectrum. One way to achieve 
this is to analyze the quasi-particle spectrum, which results from certain
expressions of the Virasoro characters $\chi (q)$ or their linear
combinations. Hitherto this analysis was mainly performed \cite{KKMM} for
formulae of the form 
\begin{equation}
\chi (q)=q^{\rm{const}}\sum\limits_{\vec{l}}\frac{q^{\vec{l}^{\,t}A\vec{l}+%
\vec{B}\cdot \vec{l}}}{(q)_{l_{1}\ldots }(q)_{l_{r}}}\,.  \label{GRRS}
\end{equation}
Here $r$ is the rank of the related Lie algebra {\bf g}, the matrix $A$
coincides with the inverse of the Cartan matrix, $\vec{B}$ characterizes 
the super-selection sector, $(q)_{l}:=\prod_{k=1}^{l}(1-q^{k})$, and there 
may be certain restrictions on the summation over $\vec{l}$. Following 
the prescription of \cite{KKMM} one can always obtain a quasi-particle spectrum once a character admits a representation in the form of 
equation (\ref{GRRS}). 
It should be noted that such spectra can not be obtained 
form the standard form of the Virasoro characters (\ref{stand}).

In the following we will demonstrate that one also recovers Lie algebraic
structures in certain Virasoro characters or their linear combinations 
which admit the factorized form 
\begin{equation}
\frac{q^{\rm{const}}}{\left\{ 1\right\} _{1}^{-}}\left\{ x_{1};\ldots
;x_{N}\right\} _{y}^{-}\left\{ x_{1}^{\prime };\ldots ;x_{M}^{\prime
}\right\} _{y}^{+}\,,  \label{fac}
\end{equation}
where we adopt the notations of \cite{BF3} 
\[
\{x\}_{y}^{\pm }:=\prod\limits_{k=0}^{\infty }\left( 1\pm q^{x+ky}\right)
\,,\quad \left\{ x_{1};\ldots ;x_{n}\right\} _{y}^{\pm
}:=\prod\limits_{a=1}^{n}\{x_{a}\}_{y}^{\pm }\,.
\]
In many cases (see \cite{BF3,BF} for details) expressions of the 
type (\ref{fac}) 
can be rewritten in the form (\ref{GRRS}), but now $A$ is entirely absent or, 
at most, is a diagonal matrix. There are no 
restrictions on the summation over $\vec{l}$, and we allow terms of the 
form $(q^{y})_{l}$ in the denominator (which may be regarded as an anionic 
feature \cite{BF}).

Unlike the conventional form for the Virasoro characters (\ref{stand}),
formulae (\ref{GRRS}) and (\ref{fac})  allow to extract the leading order
behaviour in the limit $q\rightarrow 1^{-}$ by means of a saddle point
analysis, see e.g.  \cite{KKMM,BF}. For a slightly generalized version
of (\ref{GRRS}), in the sense that all $(q)_{l}$  are replaced by 
$(q^{y})_{l}$, this analysis leads  to
\begin{equation} \label{saddle}
 z_{i}^y = \prod\limits_{j=1}^{r}(1-z_{j})^{(A_{ij}+A_{ji})},\quad 
 c_{\rm{eff}}=\frac{6}{y\pi ^{2}}\sum\limits_{i=1}^{r}L(z_{i})\,.  
\end{equation}
This means solving the former set of equations for the unknown quantities 
$z_{i}$, we may compute the effective central charge thereafter by means of
the latter equation in terms of Rogers dilogarithm $L(x)$. Recall that the
effective central charge is defined as $c_{\rm{eff}}=c-24h^{\prime}$,
where $h^{\prime }$ is the lowest conformal weight occurring in the model.
There exist inequivalent solutions to  equations (\ref{saddle}) leading 
to the same
effective central charge corresponding either to the form (\ref{GRRS}) or 
(\ref{fac}). When treating these equations as formal series, such
computations give a first hint on possible candidates for characters. 

Alternatively, with regard to factorization, we can exploit the essential
fact that the blocks $\{x\}_{y}^{\pm }$ are closely related to the 
so-called quantum dilogarithm and we can easily compute their contributions 
to the effective central charge. As explained in \cite{BF3}, each block 
$(\{x\}_{y}^{-})^{\pm 1}$ and $(\left\{ x\right\} _{y}^{+})^{\pm 1}$ in
expressions of type (\ref{fac}) contributes 
\begin{equation}
\Delta c_{\rm{eff}}=\mp 
\begin{array}{c}  \frac{1}{y} \end{array}    
\,,\quad \rm{and}\quad \Delta 
 c_{\rm{eff}}=\pm 
\begin{array}{c}  \frac{1}{2y} \end{array} 
\,,  \label{cef}
\end{equation}
respectively. In the course of our argument, i.e. when we consider the
difference of the Virasoro characters, we will also need the notion of the
secondary effective central charge 
\begin{equation}
\tilde{c}=1-24h^{\prime \prime } \, ,  \label{cs}
\end{equation}
where $h^{\prime \prime }$ is the next to lowest conformal weight occurring
in the model.

\section{ADE-Structure}

Let {\bf g} be a Lie algebra of rank $r$ and $h$ be its Coxeter number.
We define the following function related to {\bf g} 
\begin{equation}
\Xi ^{{\bf g}}(\vec{x},q)=\frac{q^{\rm{const}}}{\left\{ x_{1};\ldots
;x_{r}\right\} _{\frac{h}{2}+1}^{-}}\,,  \label{222}
\end{equation}
with $\vec{x}$ obeying the condition 
\begin{equation}
x_{a}+x_{r+1-a}=h/2+1\,,\quad a=1,\ldots ,r\,,  \label{xh}
\end{equation}
and for odd $r$ we put $x_{\frac{r+1}{2}}=\frac{h}{4}+\frac{1}{2}$. Our aim
is to find  conformal models such that their characters or possibly
linear combinations  coincide with (\ref{222})
for appropriately chosen $q^{\rm{const}}$ and $\vec{x}$.
Such conformal models have quasi-particle spectra, generated by (\ref{222})
for the related sectors, with the number of different particle 
species equal to the rank $r$.

The question arises for which conformal models can we expect 
(\ref{222}) to be a character?
Exploiting (\ref{cef}), we readily find the corresponding effective 
central charge 
\begin{equation}  \label{cg}
 c_{\rm{eff}}({\bf {g}})=
 \frac{2r}{h+2}= \frac{2r^{2}}{{\rm dim}\,{\bf g}+1} \, .
\end{equation}
On the other hand, the analysis of the ultra-violet limit of the
thermodynamic Bethe ansatz \cite{KM} for the ADE related minimal 
scattering matrices of affine Toda field theory leads to the following effective central charges:

\begin{center}
\begin{tabular}{|l|c|c|c|c|c|c|}
\hline
\vphantom{ \rule{0pt}{12pt} } {\bf g} & $A_{n}^{(1)}$ & $D_{n}^{(1)}$ & 
 $E_{6}^{(1)}$ & $E_{7}^{(1)}$ & $E_{8}^{(1)}$ & $A_{2n}^{(2)}$ \\[1pt] \hline \vphantom{\rule{0pt}{11pt}} c$_{\rm{eff}}$ \, & $\frac{2n}{n+3}$ 
 & $1$ & $\frac{6}{7}$ & $\frac{7}{10}$ & $\frac{1}{2}$ & 
 $\frac{2n}{2n+3}$ \\ [2pt] \hline
\end{tabular}
\end{center}
\vspace*{1mm} 
{\small {\ Table 1. Effective central charges for affine Toda field 
theories. } }
\vspace*{0.5mm} 

Thus, we see that, upon substitution of the related Lie algebraic
quantities\footnote{For a twisted affine Lie algebra of type $X_N^{(k)}$
one introduces $h$ -- the Coxeter number and $h^{(k)}=kh$. We should
use $h$ in (\ref{cg}) for $A_{2n}^{(2)}$.}
of the simply laced algebras (see e.g.~\cite{Kac}), eq.~(\ref{cg}) 
recovers all the effective central charges in Table 1. 
Furthermore it turns out that for {\bf g} from this table corresponding to
minimal models or $c=1$ models we are always able to identify several 
$\Xi^{{\bf g}}(\vec{x},q)$ with single Virasoro characters or specific
linear combinations of them.

In addition, there exist characters which exhibit even
stronger Lie algebraic features. They are given by (\ref{222}) with the
values of $x_{a}$ chosen as follows (which is a particular case of 
(\ref{xh})) 
\begin{equation}
2x_{a}-1=e_{a}\,,\quad a=1,\ldots ,r\,,  \label{xe}
\end{equation}
where $\{e_{a}\}$ stands for the set of the exponents of the Lie algebra 
{\bf g}. We denote this particular character as $\Xi ^{{\bf g}}(q)$. 

\subsection{Minimal Models}

The minimal models \cite{BPZ} are parameterized by a pair $(s,t)$
of co-prime positive integers and the corresponding central charge is
$c(s,t)=1-\frac{6(s-t)^{2}}{s\;t}$. Labeling the highest weights as $h_{n,m}^{s,t}=\frac{(nt-ms)^{2}-(s-t)^{2}}{4\;s\;t}$, with the 
restrictions $1\leq n\leq s-1$ and $1\leq m\leq t-1$, the usual form 
of the characters of irreducible highest weight representations reads 
\cite{ch} 
\begin{equation}
\chi_{n,m}^{s,t}(q)=\eta^{s,t}_{n,m} \!\! \sum_{k=-\infty }^{\infty }
q^{stk^{2}}\left( q^{k(nt-ms)}-q^{k(nt+ms)+nm}\right) .  \label{stand}
\end{equation}
Here we abbreviated the ubiquitous factor $\eta^{s,t}_{n,m} :=
q^{h_{n,m}^{s,t}-\frac{c(s,t)}{24}} / \left\{ 1\right\}_{1}^{-}$ by 
an analogy with the eta-function. The (secondary) effective
central charge is easy to find (see e.g.~\cite{BF3})
\begin{equation} \label{cc}
 c_{\rm{eff}}(s,t)= 1-\frac{6}{s\;t} \, , \quad 
 \tilde{c}(s,t) = 1-\frac{24}{s\;t} \, .
\end{equation}
Thus, the values of $c_{\rm{eff}}$ which are less than 1 can be
matched as follows 
\begin{eqnarray}
&& c_{\rm{eff}}(A_{1}^{(1)})= c_{\rm{eff}}(E_{8}^{(1)}) =
c_{\rm{eff}}(3,4) \, , \label{a1} \\
&&  c_{\rm{eff}}(A_{2}^{(1)})=c_{\rm{eff}}(5,6)=
 c_{\rm{eff}}(3,10) = c_{\rm{eff}}(2,15) \, , \label{a2} \\ 
&& c_{\rm{eff}}(E_{6}^{(1)})= c_{\rm{eff}}(6,7) =
 c_{\rm{eff}}(3,14) = c_{\rm{eff}}(2,21) \, , \label{e6} \\
&& c_{\rm{eff}}(E_{7}^{(1)})=c_{\rm{eff}}(4,5) \, , \\
&& c_{\rm{eff}}(A_{2n}^{(2)})=c_{\rm{eff}}(2,2n+3) \, . 
  \label{a2n} 
\end{eqnarray}
We see, that the matching is, in general, not unique. For (\ref{a2n})
it depends on $n$ -- the first non-unique representations occur for
$n=6$ and $n=9$ and coincide with (\ref{a2}) and (\ref{e6}), respectively.
Therefore we might have for instance relations between  $A_{2}^{(1)} 
\sim A_{12}^{(2)}$ and $E_{6}^{(1)} \sim A_{18}^{(2)}$. 
Some of these apparent ambiguities are easily explained as the consequence
of a symmetry property of the characters. For instance we observe
that eq.~(\ref{stand}) possesses the  symmetry:
$\chi_{\alpha n,m}^{\alpha s,t}(q)=\chi_{n,\alpha m}^{s,\alpha t}(q)$,
for instance $\chi_{2,m}^{6,5}(q)=\chi_{1,2m}^{3,10}(q)$. 

Of course eqs.~(\ref{a1})-(\ref{a2n}) are only to be understood as a first
hint on a possibility for characters in the corresponding models 
to be of the form $\Xi ^{{\bf g}}(\vec{x},q)$. 
In order to make the identifications more precise, we have to resort to 
more stringent properties of the characters. 
We shall be using  previously obtained
results \cite{BF3,PC} on representation of  characters of minimal models
in the form (\ref{fac}).
In \cite{PC} it was proven, that for $M=0$ and $x_{i}\neq x_{j}$ for 
$i\neq j$ in (\ref{fac}) the only possible factorizable single characters are 
\begin{eqnarray}
\chi_{n,m}^{2n,t}(q) &=&\eta^{2n,t}_{n,m}\, \{nm;nt;nt-nm\}_{nt}^{-} \, ,
\label{sing1} \\
\chi_{n,m}^{3n,t}(q) &=& \eta^{3n,t}_{n,m}\, \{2nt;nm;2nt-nm\}_{2nt}^{-} 
\nonumber \\
&&\times \{2nt-2nm;2nt+2nm\}_{4nt}^{-}\, .  \label{sing2}
\end{eqnarray}
Combining now characters in a linear way, the property to be factorizable
remains still exceptional. It was argued in \cite{BF3} that 
\begin{equation} \label{pm}
 \chi_{n,m}^{s,t}(q) \pm \chi_{n,t-m}^{s,t}(q)
\end{equation}
are the only combinations of characters in the same model which have
a chance to acquire the form (\ref{fac}) with reasonably small $N$ and $M$.
The limit $q \rightarrow 1^{-}$ of the upper and lower signs in 
(\ref{pm}) is governed by $c_{\rm{eff}}$ and $\tilde{c}$, respectively,
which can be seen form their properties with respect to the S-modular
transformation \cite{BF3}.

The following factorizable combinations of type (\ref{pm}) where found 
in \cite{BF3} 
\begin{eqnarray}
\chi_{n,m}^{3n,t}(q) &\pm &\chi_{2n,m}^{3n,t}(q) =\eta^{3n,t}_{n,m}
\,\left\{ nm;nt-nm\right\} _{nt}^{-}  \nonumber \\
&&\times \begin{array}{c} \left\{ \frac{nt}{2}\right\}_{\frac{nt}{2}}^{-} 
\left\{ \frac{nt-2nm}{4}; \frac{nt+2nm}{4}\right\}_{\frac{nt}{2}}^{\pm} 
\end{array}  ,  \label{f3} \\
\chi_{n,m}^{4n,t}(q) &-&\chi_{3n,m}^{4n,t}(q)=\eta^{4n,t}_{n,m}\, 
\begin{array}{c} \left\{ \frac{nt}{2};\frac{nt}{2}-nm;nm \right\}
_{\frac{nt}{2}}^{-} \end{array} \, ,  \label{f4-} \\
\chi_{n,m}^{4n,t}(q) &+&\chi_{3n,m}^{4n,t}(q)=\eta^{4n,t}_{n,m} \left\{
nm;nt;nt-nm\right\} _{nt}^{-}  \nonumber \\
&&\times \begin{array}{c} \left\{ \frac{nt}{2}-nm;\frac{nt}{2}+nm;
 \frac{nt}{2}\right\}_{nt}^{+} \end{array}\, ,  \label{f4+} \\
\chi_{n,m}^{6n,t}(q) &-&\chi_{5n,m}^{6n,t}(q)=\eta^{6n,t}_{n,m}\, \left\{
nt;nm;nt-nm\right\}_{nt}^{-}  \nonumber \\
&&\times \left\{ nt-2nm;nt+2nm\right\}_{2nt}^{-}\,.  \label{f6}
\end{eqnarray}
For the differences of type (\ref{pm}) it was proven that besides the presented formulae no other combinations are possible to factorize in 
the form (\ref{fac}) with $M=0$ and $x_{i}\neq x_{j}$ for $i\neq j$.

Remarkably, as we demonstrate in Table 2, all  identifications presented 
in (\ref{a1})-(\ref{a2n}) can be realized in terms of characters with 
the help of (\ref{sing1})-(\ref{f6}). In particular, we identify 
$\Xi^{{\bf g}}(q)$ with the following characters
\begin{eqnarray}
&& {\Xi}^{A_{1}^{(1)}}(q)=\chi_{1,2}^{3,4}(q) \, , \quad
 {\Xi}^{A_{2n}^{(2)}}(q)=\chi_{1,n+1}^{2,2n+3}(q) \, , \nonumber \\  
&& {\Xi}^{A_{2}^{(1)}}(q)=\chi_{1,2}^{5,6}(q)+\chi_{1,4}^{5,6}(q)=
\chi_{1,2}^{3,10}(q)+\chi_{1,8}^{3,10}(q) \, , \nonumber \\ 
&& {\Xi}^{E_{6}^{(1)}}(q)=\chi_{2,1}^{6,7}(q)+\chi_{2,6}^{6,7}(q) =
\chi_{1,2}^{3,14}(q)+\chi_{1,12}^{3,14}(q) \, , \nonumber \\ 
&& {\Xi}^{E_{7}^{(1)}}(q)=\chi_{2,1}^{4,5}(q) \, , \quad
 {\Xi}^{E_{8}^{(1)}}(q)=\chi_{1,3}^{3,4}(q) \, . \nonumber
\end{eqnarray}
Since these identifications hint on the connection with massive models,
i.e. affine Toda field theories, it is somewhat surprising that  
also the non-simply laced algebras
$G_2$ and $F_4$ occur in Table 2.\footnote{We thank W.~Eholzer for
pointing out to us that our formulae in \cite{BF} may include
${\Xi}^{G_2}(q)= \chi_{1,2}^{3,4}(q)$ 
and ${\Xi}^{F_4}(q) =\chi_{1,2}^{2,7}(q) $ as well.} 
No connection is known
 between $G_2-$ and $F_4-$affine Toda models and $(3,4)$ and $(2,7)
$ minimal models, respectively.\footnote{Notice also that a substitution 
of the inverse Cartan matrix 
for $G_2$ into (\ref{GRRS}) yields in the standard $q\rightarrow 1^-$
analysis $c_{\rm eff}\approx 0.453$.}
At present it seems to be just an intriguing coincidence. Replacing
in (\ref{222}) the Coxeter number by the dual Coxeter number would exclude
the non-simply laced cases without altering the formulae for the 
simply laced ones. 

It is interesting to notice that, as seen from Table 2, the differences 
of type (\ref{f6}) can also be of the form 
\begin{equation}
\frac{q^{\rm{const}}}{\left\{ x_{1};\ldots;x_{r}\right\}_{b}^{-}}.
\label{fdiff}
\end{equation}
For instance, $\chi_{1,1}^{5,6}(q) - \chi_{1,5}^{5,6}(q)$ corresponds to
$b=2h+4$, $x_a=2(e_a+1)$, and $\chi_{1,2}^{6,7}(q) - \chi_{1,5}^{6,7}(q)$
corresponds to $b=h+2$, $x_a=e_a+1$, where $h$ and $\{ e_a \}$ are the
Coxeter number and exponents of $A_2$ and $E_6$, respectively.

\begin{center}
\begin{tabular}{|c|c|c|}
\hline
{\bf g} & $\rm{sectors}$ & $\left\{ x_{1};\ldots ;x_{r}\right\}_y^-$ \\
\hline
$E_{8}$ & $\chi _{1,1}^{3,4}$ & $\{2;3;4;5;11;12;13;14\}_{16}^{-}$ \\ 
\hline
& $\chi_{1,2}^{3,4}$ & $\{1;3;5;7;9;11;13;15\}_{16}^{-}$  \\ \hline
& $\chi_{1,3}^{3,4}$ & $\{1;4;6;7;9;10;12;15\}_{16}^{-}$  \\ \hline
$A_{1}$ & $\chi _{1,2}^{3,4}$ & $\{1\}_{2}^{-}$  \\ \hline
$E_{7}$ & $\chi _{2,1}^{4,5}$ & $\{1;3;4;5;6;7;9\}_{10}^{-}$  \\ \hline
& $\chi_{2,2}^{4,5}$ & $\{1;2;3;5;7;8;9\}_{10}^{-}$   \\ \hline
$A_{2}$ & $\chi _{1,2}^{5,6} + \chi _{1,4}^{5,6}$ & 
 $\left\{ { 1;\frac{3}{2} } \right\} _{5/2}^{-}$  \\ \hline
& $\chi_{2,2}^{5,6} + \chi _{2,4}^{5,6}$ & 
 $\left\{ { \frac{1}{2} ;2 } \right\}_{5/2}^{-}$   \\ \hline
& $\chi_{1,1}^{5,6} - \chi_{1,5}^{5,6}$ & $\left\{ 2;8\right\}_{10}^- $ 
\\ \hline
& $\chi_{2,1}^{5,6} - \chi_{2,5}^{5,6}$ & $\left\{ 4;6 \right\}_{10}^- $ 
\\ \hline
$E_{6}$ & $\chi_{2,1}^{6,7} + \chi _{4,1}^{6,7}$ & 
 $\{1;{\bf \frac{5}{2}}; 3;4;{\bf \frac{9}{2}}; 6\}_{7}^-$ \\ \hline
& $\chi_{2,2}^{6,7} + \chi_{4,2}^{6,7}$ & 
 $\{1;{\bf \frac{3}{2}};2;5;{\bf \frac{11}{2}};6\}_{7}^-$  \\  \hline
& $\chi_{2,3}^{6,7} + \chi_{4,3}^{6,7}$ & 
 $\{{\bf \frac{1}{2}};2;3;4;5;{\bf \frac{13}{2} }\}_{7}^-$  \\ \hline
& $\chi_{1,1}^{6,7}-\chi_{1,6}^{6,7}$ & $\{2;3;4;10;11;12\}_{14}^-$
  \\ \hline
& $\chi_{1,2}^{6,7}-\chi_{1,5}^{6,7}$ & $\{1;4;6;8;10;13\}_{14}^{-}$
\\ \hline
& $\chi_{1,3}^{6,7}-\chi_{1,4}^{6,7}$ & $\{2;5;6;8;9;12\}_{14}^{-}$
\\ \hline
$G_2$ & $\chi_{1,2}^{3,4}$ & $\{1;3\}_{4}^-$   \\ \hline
$F_{4}$ & $\chi _{1,1}^{2,7}$ & $\{2;3;4;5\}_{7}^{-}$   \\ \hline
& $\chi_{1,2}^{2,7}$ & $\{1;3;4;6\}_{7}^{-}$  \\ \hline
& $\chi_{1,3}^{2,7}$ & $\{1;2;5;6\}_{7}^{-}$   \\ \hline
\end{tabular}
\end{center}

\vspace*{1mm}

\noindent
{\small  Table 2. Representation of characters in the form 
of  (\ref{222}) and for differences of the type (\ref{f6}) in the
form (\ref{fdiff}). The replacement of the blocks $\{x\}^-$  typed in bold 
by  $\{x\}^+$  yields the corresponding differences
 of characters.} 

\vspace*{0.5mm}

Eq.~(\ref{222}) does not exhaust all manifest Lie algebraic functions 
in which combinations of characters of the type (\ref{pm}) 
can be represented. For instance, the following representation 
\begin{equation}
\bar{\Xi} ^{{\bf g}}(\vec{w},q)= q^{\rm{const}} 
 \frac{ \left\{ \frac{h+2}{8} \right\}_{\frac{h+2}{4}}^+ }{
 \left\{ w_{1};\ldots;w_{r-1}\right\}_{\frac{h}{2}+1}^{-}}\, \label{33}
\end{equation}
with $\vec{w}$ obeying the condition $w_{a}+w_{r-a}=h/2+1$
also occurs (see Table 3).

\begin{center}
\begin{tabular}{|c|c|c|c|}
\hline
{\bf g}  & $\rm{sectors}$ & $\left\{ w_{1};\ldots ;w_{r-1}
 \right\}^-_{y}$ \\ \hline
$A_{1}$  & $\chi_{1,1}^{3,4} + \chi _{1,3}^{3,4}$ & 1 \\ \hline
$E_{7}$  &$\chi_{1,1}^{4,5} + \chi _{1,4}^{4,5}$ & 
 $ \{ {\bf \frac 32};2; {\bf \frac 72;\frac{13}{2}}; 8; {\bf 
 \frac{17}{2} } \}_{10}^{-}  $ \\ \hline
&  $\chi_{1,2}^{4,5} + \chi _{1,3}^{4,5}$ & 
 $ \{{\bf \frac 12};4;{\bf  \frac 92};{\bf \frac{11}{2}};6;
{\bf \frac{19}{2}} \}_{10}^{-} $ 
   \\ \hline
\end{tabular}

\vspace*{1mm}

\noindent
\end{center}
{\small  Table 3. Representation of characters in the form 
of  (\ref{33}). For the arguments in bold the same convention applies
as in table 2, including the numerator.}

\subsection{Compactified free Boson}

In general the Fock space of a free boson may simply be constructed from 
a Heisenberg algebra and the corresponding Virasoro central charge 
equals $c=1$. The character of the Heisenberg module ($\hat{U}(1)$-Kac-Moody) 
is simply the inverse of the $\eta$-function, 
$q^{-1/24}/\left\{ 1\right\}_{1}^{-}$. 
When compactifying the boson on a circle of rational square radius $R=\sqrt{2s/t}$ one can associate highest weight representations of a $\hat{U}(1)_{k}$-Kac-Moody algebra to this theory. 
The $\hat{U}(1)_{k}$-algebra  has an integer level, which is 
$k=st$. The corresponding characters read \cite{Kac}
\begin{equation}
 \hat{\chi}_{m}^{k}(q)= \hat{\eta}_{m}^{k}\sum_{l=-\infty }^{\infty} 
 q^{kl^{2}+ml} = \hat{\eta}_{m}^{k}\{2k\}_{2k}^{-}
 \left\{ k-m;k+m\right\}_{2k}^{+}.  \label{chD}
\end{equation}
where we denoted $\hat{\eta}_{m}^{k}:=q^{h_{m}^{k}-\frac{1}{24}}/\left\{
1\right\} _{1}^{-}$. The highest weight may take on the values 
$h_{m}^{k}=\frac{m^{2}}{4k}$ with $m=0,1,\ldots ,k-1$. 

{}As Table 1 indicates, the  $D_n$-affine Toda models are related 
to compactified bosons. In order to
recover the $D_n$-structure at the conformal level, we shall find
realizations of $\Xi^{D_{n}}(q)$ in terms of (\ref{chD}). Similar to the
case of minimal models, it will be helpful to study factorization
of (combinations of) the Kac-Moody characters.

First, choosing the constant in (\ref{222}) as $h_{n/2}^{n}-
\begin{array}{c}  \frac{1}{24} \end{array} $ 
we identify for even $n$ 
\begin{eqnarray}
 \Xi^{D_{n}}(q)&=&\hat{\eta}_{n/2}^{n} \frac{\{ n\}_{n}^{-}}{
\left\{ \frac{n}{2}\right\}_{n}^{-}}= \hat{\eta}_{n/2}^{n}
\{ n\}_{n}^{-}  \begin{array}{c} \left\{ \frac n2 
 \right\}_{\frac n2 }^{+} \end{array} \nonumber \\
&=& \hat{\eta}_{n/2}^{n} \{ 2n\}_{2n}^{-} \begin{array}{c} \left\{
 \frac n2 \right\}_{n}^{+} \end{array} =
 \hat{\chi}_{n/2}^{n}(q)\,. \label{D1}
\end{eqnarray}
Formally this expression also holds for odd $n$, albeit in this case the
right hand side may not be interpreted as the character related to a
compactified boson. Therefore we need another way  to construct
$\Xi^{D_{n}}(q)$ in case $n$ is odd. For this purpose we consider the
combinations $\hat{\chi}_{m}^{n}(q) \pm\hat{\chi}_{n-m}^{n}(q)$
which are analogues of (\ref{pm}). In particular, the $q\rightarrow 1^-$
limit of these sums and differences is governed by $c_{\rm eff}=c=1$
and $\tilde{c}$. According to (\ref{cs}) and the possible values for 
the highest weights we have  $\tilde{c}=1-6/n$.

Exploiting the identity (2.31) obtained in \cite{BF3}, we find 
(for $0 < m < n/2 $) 
\begin{eqnarray}  \label{Dpm}
&& \hat{\chi}_{m}^{n}(q) \pm\hat{\chi}_{n-m}^{n}(q)= \hat{\eta}_{m}^{n}
\begin{array}{c}
\left\{ \frac{n}{2}\right\}_{\frac{n}{2}}^{-} \left\{ \frac{n}{4} - 
\frac{m}{2}; \frac{n}{4} + \frac{m}{2} \right\}_{\frac{n}{2}}^{\pm} 
\end{array} \, .
\end{eqnarray}
The counting, based on (\ref{cef}), gives the expected values of $c$
and $\tilde{c}$. Notice that for the upper sign 
the r.h.s.~can be identified as the product side of (\ref{chD}): 
\begin{equation}  \label{id}
\hat{\chi}_{m}^{n}(q) + \hat{\chi}_{n-m}^{n}(q)= 
\hat{\chi}_{m/2}^{n/4}(q) \, .
\end{equation}
Comparison with (\ref{D1}) then yields a formula for $\Xi^{D_{n}}(q)$ 
valid for both odd and even $n$
\begin{equation}  \label{D2}
\Xi^{D_{n}}(q)=\hat{\chi}_{n}^{4n}(q) + \hat{\chi}_{3n}^{4n}(q) \, .
\end{equation}

Finally, it is interesting to observe that, employing (\ref{sing1})-(\ref{f3}), we may express (\ref{chD}) and (\ref{Dpm}) entirely in 
terms of the minimal Virasoro characters: 
\begin{eqnarray}
\hat{\chi}_{m}^{n}(q) &=& \chi_{1,m}^{3,n}(q) \, \frac{ 
\chi_{1,n}^{2,3n}(q) }{ \chi_{1,m}^{2,n}(q) } \, , \label{Dm1} \\
\hat{\chi}_{m}^{n}(q) \pm \hat{\chi}_{n-m}^{n}(q) &=& \left(
\chi_{1,m}^{3,n}(q) \pm \chi_{2,m}^{3,n}(q) \right) \frac{
\chi_{1,n}^{2,3n}(q) }{ \chi_{1,m}^{2,n}(q) } \, . \label{Dm2}
\end{eqnarray}
Here $n=6l\pm 1$, $l\in N$ if we really regard all components on the 
r.h.s. as characters 
of irreducible Virasoro representations. This restriction can be omitted 
if we regard (\ref{sing1})-(\ref{f3}) just as formal series. With regard 
to the central charge eqs.~(\ref{Dm1})-(\ref{Dm2}) imply
\begin{equation}
 c_{\rm{eff}}(D_{n}^{(1)})= c_{\rm{eff}}(3,n) + c_{\rm{eff}}(2,3n) - 
 c_{\rm{eff}}(2,n) \, .
\end{equation}
Thus, the connection with minimal models is more subtle than
one would expect at first sight from a simple matching of the 
central charges, e.g.~$c_{\rm{eff}}(D_{n}^{(1)})= 2 c_{\rm{eff}}(3,4)$. 

\subsection{Parafermions}

The $A_n^{(1)}$-series of affine Toda theories is known to be related 
 in the ultra-violet limit (see e.g. \cite{KM}) to the 
$Z_{n+1}$-parafermions \cite{pf}. The corresponding central charge,
$c(k)=2(k-1)/(k+2)$ and characters may be obtained from the 
$\hat{SU}(2)_{k-1}/\hat{U}(1)$-coset, where $k=n+1$. Introducing the quantity  $\Delta_{j,m}^{k}=j(j+1)/(k+2)-m^{2}/k$ the characters of the highest weight representation, which appear as branching
functions in the coset, acquire the form \cite{KP} 
\begin{eqnarray}
 && \tilde{\chi}_{j,m}^{k}(q) = \frac{\tilde{\eta}_{j,m}^{k} }{ 
 \{ 1 \}_{1}^{-} }\sum_{r,s=0}^{\infty}(-1)^{r+s}q^{rs(k+1)+  
 \frac{r(r+1)}{2}+ \frac{s(s+1)}{2}} \label{br} \\
 && \times \left( q^{r(j+m)+s(j-m)}-
  q^{r(k+1-j-m)+s(k+1-j+m)+k+1-2j}\right) , \nonumber
\end{eqnarray}
where $\tilde{\eta}_{j,m}^{k}:=q^{\Delta _{j,m}^{k}-\frac{c(k)}{24}} / 
 \{ 1\}_{1}^{-} $. The labels are restricted as $-j\leq m\leq k-j$,
 $0\leq j\leq k/2$ and $(j-m)\in Z$. 
In particular the $\tilde{\chi}_{0,m}^{k}(q)$
are  the characters of the parafermionic  currents
$\psi^k_m $. The characters possess the symmetries 
\begin{equation}
\tilde{\chi}_{j,m}^{k}(q)=\tilde{\chi}_{j,-m}^{k}(q)=
 \tilde{\chi}_{k/2-j,k/2-m}^{k}(q) \, . \label{brs}
\end{equation}

{}From our observations made above for characters of the ADE-related
conformal models one may expect that expressions (\ref{br}) exhibit
$A_n$-type structures (e.g. posses $n$ quasi particles and moreover 
acquire the form of the type 
$\Xi^{A_{n}}(q)$) in some of the cases when they admit a
factorized form. We shall now discuss this issue in detail for
several of the lowest ranks. 

As we have seen above, factorization of linear combinations
of characters occurs usually only for the specific type of combinations.
Now the analogue of (\ref{pm}) is 
$\tilde{\chi}_{j,m}^{k}(q) \pm \tilde{\chi}_{j,2k-m}^{k}(q)$.
One expects that the $q\rightarrow 1^-$ limit of these sums and 
differences is governed by $c_{\rm eff}=c(k)$ and $\tilde{c}$,
respectively. 
Since $h''=\Delta_{1,1}^{k}$, eq.~(\ref{cs}) yields 
$\tilde{c}=c(k)(k-6)/k$. This is confirmed by all the examples
given below.

For the parafermionic formulae (\ref{br}) we do not 
have such powerful analytical tools (analogues to the factorization
formulae (\ref{sing1})-(\ref{f6}) ) at hand as in the case of the
minimal models. Therefore, as a first step, we resort to an analysis 
with Mathematica. Typically we expand the characters up to $q^{100}$.

\noindent $\underline{ A_1 }:$  In this case there are only three 
distinct (up to the symmetries (\ref{brs})) characters and they 
can be matched with those of the (3,4) minimal model:
\begin{eqnarray}
&&   \tilde{\chi}_{0,0}^{2}(q) = \chi_{1,1}^{3,4}(q)\, , \quad
   \tilde{\chi}_{0,1}^{2}(q) = \chi_{1,3}^{3,4}(q)\, , \\
&&  \tilde{\chi}_{\frac 12, \frac 12}^{2}(q) =
   \chi_{1,2}^{3,4}(q) = \Xi^{A_{1}}(q) \, .
\end{eqnarray}
Thus, all the characters in this case are factorizable and moreover
$\Xi^{A_{1}}(q)$ is present among them.

\noindent $\underline{ A_2 }:$  There are four distinct 
characters in this case and they can be matched with those of the (5,6) 
minimal model:
\begin{eqnarray}
&&   \tilde{\chi}_{0,0}^{3}(q) = \chi_{1,1}^{5,6}(q) +
 \chi_{1,5}^{5,6}(q) \, , \quad
   \tilde{\chi}_{0,1}^{3}(q) = \chi_{1,3}^{5,6}(q)\, , \\
&&  \tilde{\chi}_{\frac 12, \frac 12}^{3}(q) =
   \chi_{2,3}^{5,6}(q) \, , \quad
\tilde{\chi}_{\frac 12, \frac 32}^{3}(q) =
 \chi_{2,1}^{5,6}(q) + \chi_{2,5}^{5,6}(q)  \, .
\end{eqnarray}
Only $\tilde{\chi}_{0,1}^{3}(q)$ and $\tilde{\chi}_{\frac 12,
 \frac 12}^{3}(q)$ are factorizable (see subsection II.A).

\noindent $\underline{ A_3 }:$  Since $c=1$, it is suggestive to try to relate
the characters to those of the compactified bosons. 
This turns out to be possible for all the characters (thus factorizability 
is guaranteed):
\begin{eqnarray}
 \tilde{\chi}_{0,1}^{4}(q) = \hat{\chi}_{6}^{12}(q) \, , && \quad
 \tilde{\chi}_{0,0}^{4}(q) + \tilde{\chi}_{0,2}^{4}(q) = 
   \hat{\chi}_{0}^{3}(q) \, ,  \label{A41} \\
 \tilde{\chi}_{1,0}^{4}(q) = \hat{\chi}_{2}^{3}(q) \, , &&\quad
 \tilde{\chi}_{1,1}^{4}(q) = \hat{\chi}_{1}^{3}(q) \, , \\
 \tilde{\chi}_{\frac 12, \frac 12}^{4}(q) = \hat{\chi}_{1}^{4}(q) \, ,&&\quad
 \tilde{\chi}_{\frac 12, \frac 32}^{4}(q) = \hat{\chi}_{3}^{4}(q) \, .
\end{eqnarray}
Furthermore, some of linear combinations can be expressed in terms
of the characters of the (3,4) minimal model (notice that $c=1$,
 $\tilde{c}=-1/2$ for $A_3$ and $c=1/2$, $\tilde{c}=-1$ for the (3,4)
minimal model):
\begin{eqnarray}
&& \tilde{\chi}_{0,0}^{4}(q) - \tilde{\chi}_{0,2}^{4}(q)  = 
 \left( {\chi}_{1,2}^{3,4}(q) \right)^{-1} \, , \\
&& \tilde{\chi}_{\frac 12, \frac 12}^{4}(q) \pm
 \tilde{\chi}_{\frac 12, \frac 32}^{4}(q) =    \label{A42}
 \left( {\chi}_{1,1}^{3,4}(q) \mp {\chi}_{1,3}^{3,4}(q) \right)^{-1} \, .
\end{eqnarray}

\noindent $\underline{ A_4 }:$  No characters or linear combinations factorize.

\noindent $\underline{ A_5 }:$  Several characters and combinations 
are factorizable and can be expressed via those of the (3,4) minimal
model and $D_n$, for instance
\begin{eqnarray*}
\tilde{\chi}_{\frac{3}{2},\frac{1}{2}}^{6}(q) &=& 
 \chi_{1,2}^{3,4}(q) \left(\hat{\chi}_{6}^{24}(q)-
 \hat{\chi}_{18}^{24}(q) \right) \, , \\
\tilde{\chi}_{\frac{1}{2},\frac{3}{2}}^{6}(q) &=& 
\chi_{1,2}^{3,4}(q) \left(\hat{\chi}_{8}^{24}(q)-
 \hat{\chi}_{16}^{24}(q) \right) \, , \\
\tilde{\chi}_{0,1}^{6}(q) \pm \tilde{\chi}_{0,2}^{6}(q) &=& 
( \chi_{1,1}^{3,4}(q) \pm \chi_{1,3}^{3,4}(q))
(\hat{\chi}_{9}^{24}(q) \mp \hat{\chi}_{15}^{24}(q) )  , \\
\tilde{\chi}_{1,1}^{6}(q) \pm \tilde{\chi}_{1,2}^{6}(q) &=&
( \chi_{1,1}^{3,4}(q) \pm \chi_{1,3}^{3,4}(q) )
(\hat{\chi}_{3}^{24}(q) \mp \hat{\chi}_{21}^{24}(q) )  .
\end{eqnarray*}
Also we notice that $\tilde{\chi}_{\frac 12, \frac 12}^{6}(q) 
- \tilde{\chi}_{\frac 12, \frac 52}^{6}(q) = q^{5/96}$. Such an identity
can occur only in this parafermionic model since it requires 
$\tilde{c}=0$.

\noindent $\underline{ A_6 }:$   -- no combinations factorize and the 
only factorizable single characters are 
\begin{eqnarray}
\tilde{\chi}_{1,m}^{7}(q) &=& \tilde{\eta}_{1,m}^{7} 
\frac{ \{3 \}^-_3  \{m;7-m;7 \}^-_7  } 
{ \{ 1 \}^-_1 \{3m;21-3m\}^-_{21} } \, , \quad m=1,2,3 .
\end{eqnarray}

Summarizing these data, we see that, apart from the $A_1$ case,
none of the factorizable (combinations of) characters provided by 
eq.~(\ref{br}) for $A_n$ can be identified as $\Xi^{A_{n}}(\vec{x},q)$. 
However, it is plausible to speculate that in general the
$\Xi^{A_n}(\vec{x},q)$ might be identifiable as characters of other conformal 
models having the central charge $2n/(n+3)$. This conjecture is supported 
by the $A_2$ case (see Table 2) and $A_3$ case, in which we can identify
\begin{equation}  \label{A3}
\Xi^{A_{3}}(q)= \hat{\chi}_{3}^{12}(q) + \hat{\chi}_{9}^{12}(q) \, .
\end{equation}

To conclude the discussion on factorizable parafermionic characters,
we notice an intriguing fact -- some characters in the $A_7$ case 
exhibit an $E_7$ structure  (cf. Tables 2 and 3): 
\begin{eqnarray*}
&& \tilde{\chi}_{0,1}^8 (q) + \tilde{\chi}_{0,3}^8 (q) = 
 \left( \chi_{2,1}^{4,5}(q) \right)^2 \, , \quad \\
&&\tilde{\chi}_{1,1}^8 (q) + \tilde{\chi}_{1,3}^8 (q) =
 \left( \chi_{2,2}^{4,5}(q) \right)^2 \, , \\
&& \tilde{\chi}_{0,0}^8(q) \pm 2 \tilde{\chi}_{0,2}^8(q) + 
 \tilde{\chi}_{0,4}^8(q) = \left( \chi_{1,1}^{4,5}(q) \pm
 \chi_{1,4}^{4,5}(q) \right)^2 \, , \\
&& \tilde{\chi}_{1,0}^8(q) \pm 2 \tilde{\chi}_{1,2}^8(q) + 
 \tilde{\chi}_{1,4}^8(q) = \left( \chi_{1,2}^{4,5}(q) \pm
 \chi_{1,3}^{4,5}(q) \right)^2 \, .
\end{eqnarray*}
This is the first case in which  we have to combine three characters
in order to obtain a factorized form.
A more detailed account on the factorization of $A_n$-related 
characters will be presented elsewhere.

\section{One Particle States}
The functions $\{x\}^+_y$ and $\frac{1}{\{x\}^-_y}$ can be written
as double series in $q$ with coefficients being ${\cal P}(n,m)$ (or 
${\cal Q}(n,m)$) -- the number of partitions of an integer $n \geq 0$
into $m$ distinct (or smaller than $m+1$) non-negative integers (see
e.g. \cite{And,BF3}).  

Applying this fact to a character of the type (\ref{fac}) with 
$x_i\neq x_j$, we obtain it in the form of a series 
$\chi(q)=\sum_{k=0}^{\infty} \mu_k q^k$, 
where the level $k$ admits the partitioning, $k=\sum_a \sum_{i_a} p_a^{i_a}$, 
into parts of a specific form (e.g.~(\ref{fer})
and (\ref{bos}) below). The interpretation of the $p_a^{i_a}$ as momenta of
massless particles gives rise to a quasi-particle picture (developed
originally for characters of the form (\ref{GRRS}) in \cite{KKMM}),
where a character is regarded as a partition function, 
$\chi(q)=\sum_k \mu_k e^{-\beta E_k}$. Here $q=e^{-2\pi \beta v/L}$,
with $v$ being the speed of sound, and $L$ -- the size of the system.
A quasi-particle spectrum constructed in this way is in
one-to-one correspondence to the Verma module of the corresponding
irreducible representations of the Virasoro algebra or some modules 
related to linear combinations. 
It is crucial to stress that this procedure is not applicable 
to the standard  representation of the characters  (i.e.~of the type 
(\ref{stand})) and is a very specific feature of the representations
(\ref{GRRS}) and (\ref{fac}).  Note that the modules which are of the 
form (\ref{GRRS}) do in general (if they do, they give rise to 
Rogers-Ramanujan type identities \cite{BF3}) not factorize, 
such that the spectra related 
to (\ref{fac}) do not only differ in nature from the ones obtainable from
(\ref{GRRS}), but are also related to different sectors. 

As just explained, a quasi-particle representation can be constructed
for any factorizable character of the type (\ref{fac}) provided that
$x_i\neq x_j$. For instance, the characters (\ref{D1}) related to
a compactified boson admit a representation with $(2k+1)$ particles.
However, since we are particularly interested in spectra with 
Lie algebraic features,
it is most natural to perform the quasi-particle analysis for the
characters which admit the form $\Xi^{{\bf g}}(\vec{x},q)$. In this
way we obtain the following fermionic spectrum (if we employ the
series involving ${\cal P}(n,m)$) in the units of $2\pi/L$
\begin{equation} \label{fer}
p_{a}^{i}(\vec{m})=x_{a}+ \begin{array}{c} \left(
 \frac{h}{4}+\frac{1}{2}\right) \left( 1-m_a\right) +
\left( \frac{h}{2}+1\right) N_{a}^{i} \end{array} ,
\end{equation}
or bosonic spectrum (if we use the series with ${\cal Q}(n,m)$) 
\begin{equation} \label{bos}
p_{a}^{i}=x_{a}+ \begin{array}{c} \left( 
 \frac{h}{2}+1\right) N_{a}^{i} \end{array} \, .
\end{equation}
Here $\vec{x}, \vec{m}$ and $\vec{N}$ parameterize the possible states.
In eq.~(\ref{fer}) the numbers $N_{a}^{i}$  are distinct positive 
integers such that $\sum_{i=1}^{m_a} N_{a}^{i} = N_{a}$, whereas in  
eq.~(\ref{bos}) they are arbitrary non-negative integers.
Notice that for the combination of characters the levels may be half 
integer graded, such that also the momenta take on half integer values in
this case. 
A sample spectrum is presented in Table 4 which illustrates how the
available momenta of the form (\ref{bos}) are to  be assembled in order to
represent a state at a particular level. 
\begin{center}
\begin{tabular}{|l|l|l|}
\hline
 $k$ &$\mu_k$& $p^{i}_1=1+ \frac{5i}{2}$, 
  $p^{i}_2= \frac 32+ \frac{5i}{2}$  \\ [0.5mm] \hline 
 $\frac 12$ & 0 & ---  \\ [0.5mm] \hline 
 1  & 1 & $ | p_1^0 \rangle $  \\ [0.5mm] \hline 
 $\frac 32$ & 1 & $ | p_2^0 \rangle $  \\ [0.5mm] \hline
 2 & 1 & $ | p_1^0,p_1^0 \rangle $  \\ [0.5mm] \hline
 $\frac 52$ & 1 & $ | p_1^0, p_2^0 \rangle $ \\ [0.5mm] \hline 
 3 & 2 & $ | p_1^0, p_1^0, p_1^0 \rangle $, 
$ | p_2^0,  p_2^0  \rangle $\\ [0.5mm] \hline
 $\frac 72$ & 2 & $ | p_1^0, p_1^0, p_2^0 \rangle $, $ | p_1^1 \rangle $ \\ 
 [0.5mm] \hline
 4 & 3 & $|p_1^0,p_1^0,p_1^0,p_1^0 \rangle $, $ | p_1^0, p_2^0,p_2^0 
\rangle $,
 $ | p_2^1   \rangle $  \\ [0.5mm] \hline
\end{tabular}
\end{center}

\vspace*{1mm}

\noindent
{\small  Table 4.  
Bosonic spectrum for $\chi_{1,2}^{5,6}(q) + \chi _{1,4}^{5,6}(q)$.
$k$ denotes the level and $ \mu_k $ its degeneracy. } \\[1mm]

Naturally the questions arise if we can interpret these spectra more deeply
and if we can  possibly find  alternative representations for the
related modules.  
First of all we should give a meaning to the particular combinations 
which occur in our analysis. In \cite{BF3} we provided several 
possibilities. In particular the combination
$\chi_{1,2}^{5,6}(q) + \chi_{1,4}^{5,6}(q)$ is  of interest in
the context of boundary conformal field theories,
 since this combination of characters coincides
with the partition function $Z_{A,F}$ for the critical 3-state
Potts model with boundaries \cite{Car} ($F$ denotes the free boundary
condition). It is intriguing that this combination possesses a 
manifestly Lie algebraic quasi-particle spectrum.
The combination 
$\chi_{2,2}^{5,6}(q) + \chi_{2,4}^{5,6}(q)$, which  
coincides with $Z_{BC,F}$ in the same model possesses a slightly
weaker relation to $A_2$.  

To answer the question concerning possible representations, 
we recall the fact that the fields corresponding to the highest
weight states satisfy the quantum equation of motion of Toda field
theory \cite{W}. It is therefore very suggestive to try to identify
the presented spectra in terms of the $W$-algebras \cite{LF}.  
For $\Xi ^{{\bf g}}(q)$ we  can make this more manifest.
Changing the units of the momenta to $\pi /L$, we obtain from (48) 
\begin{equation}
p_{a}^{i} =e_{a}+1+(h+2)N_{a}^{i} \, .
\end{equation}
Here $e_{a}$ belongs to the exponents of the Lie algebra. Since 
the generators of the W-algebras $W_{s+1}$ are graded by 
the exponents plus one \cite{FZ}, we may associate the following 
generators to this quasi-particle spectrum 
\begin{equation}
p_{a}^{i} \sim W_{a}\left( W_{a}W_{r-a}\right)^{N^i_a} \, .
\end{equation}
In particular, the critical 3-state Potts model with boundaries
would be related to the $W_3$-algebra.
We leave it for the future to investigate this conjecture in more
detail.

{\bf Acknowledgment:} We would like to thank W.~Eholzer for useful
discussions. A.B. is grateful to the members of the Institut
f\"ur Theoretische Physik, FU-Berlin for hospitality. A.F. is grateful to
the Deutsche Forschungsgemeinschaft (Sfb288) for partial support.

\end{document}